\documentclass[aps,superscriptaddress]{revtex4}

\usepackage{amsfonts,amsmath,amssymb,mathrsfs,bbm,bm}
\usepackage{hyperref}
\usepackage{graphicx}
\usepackage{multirow}

\begin{document}

\title{Complexity-Entropy Causality Plane as a Complexity Measure for Two-dimensional Patterns}

\author{Haroldo V. Ribeiro}\email{hvr@dfi.uem.br}
\affiliation{Departamento de F\'isica and National Institute of Science and Technology for
Complex Systems, Universidade Estadual de Maring\'a, Maring\'a, PR 87020-900, Brazil}
\author{Luciano Zunino}
\affiliation{Centro de Investigaciones \'Opticas (CONICET La Plata - CIC), C.C. 3, 1897 Gonnet, Argentina}
\affiliation{Departamento de Ciencias B\'asicas, Facultad de Ingenier\'ia, Universidad Nacional de La Plata (UNLP), 1900 La Plata, Argentina}
\author{Ervin K. Lenzi}
\author{Perseu A. Santoro}
\author{Renio S. Mendes}
\affiliation{Departamento de F\'isica and National Institute of Science and Technology for
Complex Systems, Universidade Estadual de Maring\'a, Maring\'a, PR 87020-900, Brazil}
\date{\today}

\begin{abstract}
Complexity measures are essential to understand 
complex systems and there are numerous definitions to 
analyze one-dimensional data. However, extensions of 
these approaches to two or higher-dimensional data, 
such as images, are much less common. Here, we reduce 
this gap by applying the ideas of the permutation 
entropy combined with a relative entropic index. 
We build up a numerical procedure that can be 
easily implemented to evaluate the complexity of 
two or higher-dimensional patterns. We work out this 
method in different scenarios where numerical 
experiments and empirical data were taken into 
account. Specifically, we have applied the method 
to $i)$ fractal landscapes generated  numerically 
where we compare our measures with the Hurst 
exponent; $ii)$ liquid crystal textures where 
nematic-isotropic-nematic phase transitions were 
properly identified; $iii)$ 12 characteristic 
textures of liquid crystals where the different 
values show that the method can distinguish 
different phases; $iv)$ and Ising surfaces where our 
method identified the critical temperature and also
proved to be stable.
\end{abstract}

\maketitle

\section{Introduction}
Investigations related to the so called complex systems are widely spread among different
scientific communities, ranging from physics and biology to economy and 
psychology. A considerable part of these works deals with empirical data
aiming to extract patterns, regularities or laws that rule the dynamics of the 
system. In this direction, the concept of complexity measures often
emerges. Complexity measures can compare empirical data such as time
series and classify them in somewhere between regular, chaotic or random~\cite{Rosso},
while other complexity measures can differentiate between degrees of correlations~\cite{Rosso-Z}.
Examples of these measures include algorithmic complexity~\cite{Kolmogorov},
entropies~\cite{Shannon}, relative entropies~\cite{Kullback}, fractal dimensions~\cite{Mandelbrot}, 
and Lyapunov exponents~\cite{Lyapunov}. These seminal works are still motivating new definitions,
and today there are numerous definitions of complexity, which have been successful 
applied to different areas such as medicine~\cite{Maes,Khader},
ecology~\cite{Parrott,Mendes,Parrott2,Jost}, astrophysics~\cite{Schwarz,Consolini,Lovallo}, and music~\cite{Boon,Su}.

It is surprising that this large number of complexity measures is mainly focused on one-dimensional
data, while much less attention has been paid to two and higher-dimensional structures
such as images. Naturally, there are few exceptions such as the work of Grassberger~\cite{Grassberger}
and more recent Refs.~\cite{Andrienko,Feldman,Cai}, though some of the authors of these papers agree
that a higher-dimensional approach still represents an open and subtle problem.
Furthermore, as it was stated by Bandt and Pompe~\cite{Bandt}, most of the complexity measures
depend on specific algorithms or recipes for processing the data which may also depend on tuning 
parameters. As a direct consequence, there are huge difficulties for reproducing previous results
without the knowledge of details of the methods. 

Bandt and Pompe not only raised this problem, but they also proposed an alternative method that tries to overcome 
the previous problems, introducing what they call \textit{permutation entropy} --- a \textit{natural} complexity measure for time series. 
There are many recent applications of this new technique that confirm its usefulness~
\cite{Ouyang,Li,Nicolaou,Masoller,Barreiro,Canovas,Nicoletta,Ribeiro}. In particular, Rosso et al.~\cite{Rosso}
have successful applied the Bandt and Pompe ideas together with a relative entropic measure~\cite{Lamberti} 
to differentiate chaotic time series from stochastic ones. They have constructed a diagram, which was first proposed by L\'opez-Ruiz et al.~\cite{Lopez-Ruiz}, (called as complexity-entropy causality plane) by plotting the relative entropic measure versus the permutation entropy. Intriguingly, chaotic and stochastic series are located in different regions of this representation space.

Here, we show that the complexity-entropy causality plane can be extended for higher-dimensional
patterns. We apply this new approach in different scenarios related to two-dimensional structures
and the results indicate that the method is very promising for distinguishing between 
two-dimensional patterns. The following sections are organized as follows. Section II is
devoted to review briefly the properties of the permutation information-theory-derived quantifiers
and the complexity-entropy causality plane, and also to define an appropriate way to generalize these 
definitions to higher-dimensional data. In Section III, we work out
several applications based on numerical and empirical data. Section IV presents
a summary of our results. 

\section{Methods}
The ingenious idea of Bandt and Pompe~\cite{Bandt} was to define a measure that may be easily applied
to any type of time series. The method lies on associating symbolic sequences to the segments of the 
time series based on the existence of local order, and next, by using probability distribution associated 
to these symbols, to estimate the complexity quantifier. 
For purpose of definition, let us consider a time series $\{x_t\}_{t=1,\dots,n}$ composed by $n$ elements and
also $d$-dimensional vectors ($d>1$) defined by
\[
(s)\mapsto (x_{s-(d-1)},x_{s-(d-2)},\dots,x_{s-1},x_{s})\;,
\]
where $s=d,d+1,\dots,n$. Next, for all the $(n-d+1)$ vectors, we evaluate the permutations 
\mbox{$\pi=(r_0,r_1,\dots,r_{d-1})$} of $(0,1,\dots,d-1)$ defined by 
$x_{s-r_{d-1}}\leq x_{s-r_{d-2}}\leq \dots \leq x_{s-r_{1}} \leq x_{s-r_{0}}$.
The $d\, !$ possible permutations of $\pi$ will be the accessible states of the system,
and for each state we estimate the ordinal pattern probability given by
\[
p(\pi) = \frac{\#\{s|s\leq n-d+1;~ (s) ~\text{has type}~ \pi \}}{n-d+1}\;,
\]
where the symbol $\#$ stands for the number of occurrences of the permutation $\pi$.
Now, we can apply the ordinal patterns probability distribution, $P=\{p(\pi)\}$,
to estimate a complexity measure based on some entropic formulation. 

\begin{figure}[!ht]
\centering
\includegraphics[scale=0.98]{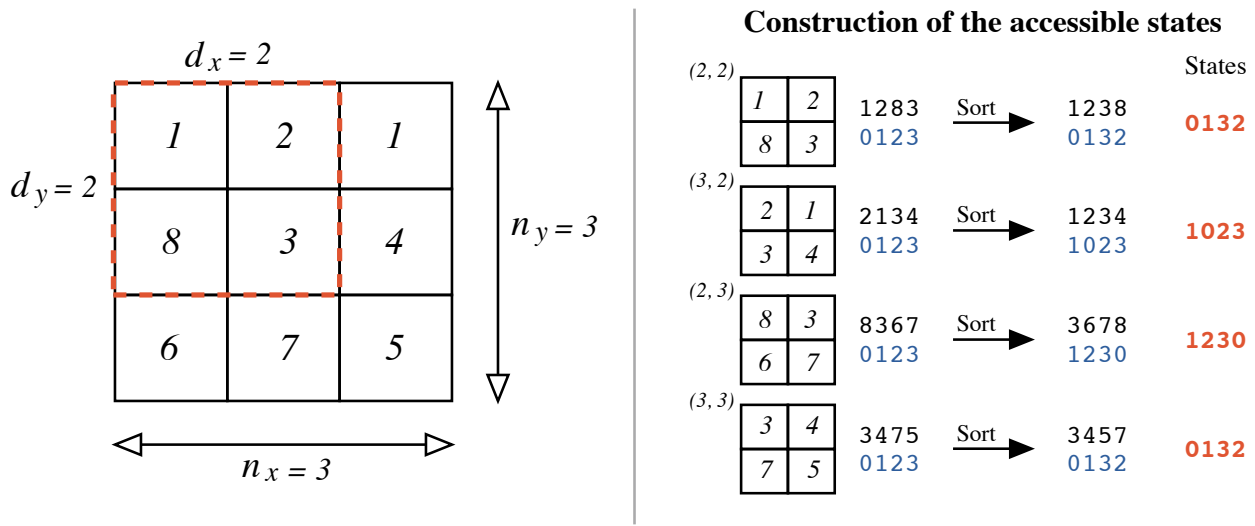}
\caption{{\textbf{Schematic representation of the construction of the accessible states}. In this example we have
a $3\times 3$ array (left panel) and we choose the embedding dimensions $d_x=2$ and $d_y=2$. In the right panel
we illustrate the construction of the states. We first obtain the sub-matrix corresponding to $s_x=2$ and $s_y=2$ that
have as elements $(1,2,8,3)$ and, after sorting, this sub-matrix leads to the state ``0132''. We thus move to next
sub-matrix $s_x=3$ and $s_y=2$ which have the elements $(2,1,3,4)$ and that, after sorting, leads to the state ``1023''.
The last two remaining matrices lead to the states ``1230'' and ``0132''. Finally, we estimate the probabilities 
$p(\pi)$, that are, $p(\text{``0132''})=2/4=0.5$, $p(\text{``1023''})=1/4=0.25$ and $p(\text{``1230''})=1/4=0.25$ which
are then used in the equations (\ref{eq:entropy}) and (\ref{eq:rosso}), leading to $H\approx0.33$ and $C\approx0.27$.
}}\label{fig:ext}
\end{figure}

Before advancing, we note that the previous method may be extended
to higher-dimensional data structures such as images. In order to do this,
we consider that the system is now represented by a
two-dimensional array $\{y_{i}^{j}\}_{i=1,\dots,n_x}^{j=1,\dots,n_y}$ of size $n_x \times n_y$. In
analogy to the vector $(s)$, we define $d_x\times d_y$ matrices ($d_x,d_y>1$) given by
\[
\begin{footnotesize}
(s_x,s_y) \mapsto
\left(
\begin{array}{ccccc}
y_{s_x-(d_x-1)}^{s_y-(d_y-1)} & y_{s_x-(d_x-2)}^{s_y-(d_y-1)} & \dots & y_{s_x-1}^{s_y-(d_y-1)} & y_{s_x}^{s_y-(d_y-1)}\\
y_{s_x-(d_x-1)}^{s_y-(d_y-2)} & y_{s_x-(d_x-2)}^{s_y-(d_y-2)} & \dots & y_{s_x-1}^{s_y-(d_y-2)} & y_{s_x}^{s_y-(d_y-2)}\\
\vdots & \vdots & \ddots & \vdots & \vdots\\
y_{s_x-(d_x-1)}^{s_y-1} & y_{s_x-(d_x-2)}^{s_y-1} & \dots & y_{s_x-1}^{s_y-1} & y_{s_x}^{s_y-1}\\
y_{s_x-(d_x-1)}^{s_y} & y_{s_x-(d_x-2)}^{s_y} & \dots & y_{s_x-1}^{s_y} & y_{s_x}^{s_y}\\
\end{array} \right)\,,
\end{footnotesize}
\]
where $s_x=d_x,d_x+1,\dots,n_x$ and $s_y=d_y,d_y+1,\dots,n_y$. Next, for all these $(n_x-d_x+1)(n_y-d_y+1)$
matrices, we evaluate the permutations 
$\pi=[(r_0,u_0),(r_1,u_0),\dots,(r_{d_x-1},u_0), \dots,(r_0,u_{d_y-1}),$ $(r_1,u_{d_y-1}),\dots,(r_{d_x-1},u_{d_y-1})]$ of 
$(0,1,\dots,d_x\,d_y-1)$ defined by 
$y_{s_x-r_{d_x-1}}^{s_y-u_{d_y-1}}\leq y_{s_x-r_{d-2}}^{s_y-u_{d_y-1}}\leq \dots \leq y_{s_x-r_{1}}^{s_y-u_{d_y-1}} \leq y_{s_x-r_{0}}^{s_y-u_{d_y-1}} \leq \dots \leq
y_{s_x-r_{d_x-1}}^{s_y-u_0}\leq y_{s_x-r_{d-2}}^{s_y-u_0}\leq \dots \leq y_{s_x-r_{1}}^{s_y-u_0} \leq y_{s_x-r_{0}}^{s_y-u_0}$. The system can now access $(d_x\,d_y)!$ states for which we calculate the probability distribution $P=\{p(\pi)\}$ through the relative frequencies given by
\[
p(\pi) = \frac{\#\{(s_x,s_y)|s_x\leq n_x-d_x+1~\text{and}~s_y\leq n_y-d_y+1;~ (s_x,s_y) ~\text{has type}~ \pi \}}
{(n_x-d_x+1)(n_y-d_y+1)}\;.
\]
For easier understanding, we illustrate this procedure for a small array in Fig.~\ref{fig:ext}.

Naturally, the order procedure that defines the permutation $\pi$ is no longer unique as in 
the one-dimensional case. For instance, instead of ordering the elements of $(s_x,s_y)$ row-by-row,
we could also order column-by-column. However, these other definitions will only change the 
``name'' of the states in such a way that the set $P=\{p(\pi)\}$ will remain unchanged. 
Thus, there is no lost of generalization in assuming a given order recipe for defining $\pi$. 

We note that this procedure is straightforward generalized to accomplish higher-dimensional structures
(e.g., the volumetric brain images obtained via functional magnetic resonance imaging), and that it 
recovers the one-dimensional case by setting $n_y=1$ and $d_y=1$. Here, for simplicity, we focus our analysis on 
two-dimensional structures.

The parameters $d_x$ and $d_y$ (known as embedding dimensions) play an important role in 
the estimation of the permutation probability distribution $P$, since they determine the 
number of accessible states. In the one-dimensional case, it is usual to choose $d\,!\ll n$ in order to obtain 
reliable statistics in the one-dimensional case (for practical purposes, 
Bandt and Pompe recommend $d=3,\dots,7$~\cite{Bandt}). For the two-dimensional case a similar relationship 
must hold, i.e., $(d_x d_y)!\ll n_x n_y$. 

To go further, we need to rewrite the entropic measures used in Refs.~\cite{Bandt,Rosso}. 
The first one is called normalized permutation entropy~\cite{Bandt} and it is obtained by applying
the Shannon's entropy to the probabilities $P=\{p(\pi)\}$, i.e., 
\begin{equation}\label{eq:entropy}
H[P]=\frac{S[P]}{S_{\text{max}}} \;,
\end{equation}
where $S[P]=-\sum p(\pi) \log p(\pi)$ and $S_{\text{max}}=\log [(d_x d_y)!]$. The value of $S_{\text{max}}$ is obtained by considering
all the $(d_x d_y)!$ accessible states to be equiprobable, i.e., $P=P_e=1/(d_x d_y)!$. By definition, 
$0 \leq H[P] \leq 1$, where the upper bound occurs for a completely random array. We expect 
$H[P]<1$ for arrays that exhibit some kind of correlated dynamics.

The other measure~\cite{Rosso} is defined by 
\begin{equation}\label{eq:rosso}
C[P]=Q[P,P_e]\,H[P]\,,
\end{equation}
where $Q[P,P_e]$ is a relative entropic metric between the empirical ordinal probability $P=\{p(\pi)\}$ and
the equiprobable state $P_e=1/(d_x d_y)!$. The quantity $Q[P,P_e]$ is known as disequilibrium
and it is defined in terms of the Jensen-Shannon divergence~\cite{Grosse} (or also
in terms of a symmetrized Kullback-Leibler divergence~\cite{Lin}) and can be written as
\begin{equation}
Q[P,P_e] = \frac{S[(P+P_e)/2] - S[P]/2 - S[P_e]/2}{Q_{\text{max}}}\,,
\end{equation}
where 
\[
Q_{\text{max}} = -\frac{1}{2}\left\{ \frac{(d_x d_y)!+1}{(d_x d_y)!} \log[(d_x d_y)!+1] -
2 \log[2 (d_x d_y)!] + \log[(d_x d_y)!] \right\}\,
\]
is the maximum possible value of $Q[P,P_e]$, obtained when one of the components of $P$ is equal to one and all the other vanish. 

The disequilibrium $C$ quantifies the degree of correlational structures
providing important additional information that may not be carried only by the permutation entropy. 
In addition, for a given $H[P]$ value there exists a range of possible values for $C[P]$~\cite{Martin2}. 
This is the main reason why Rosso et al.~\cite{Rosso} proposed to employ a diagram of $C[P]$ versus $H[P]$ as a diagnostic tool, 
building up the complexity-entropy causality plane.

\section{Applications}
In the following, we will calculate the diagram of $C[P]$ versus $H[P]$ 
to measure the complexity and to distinguish among different two-dimensional 
patterns.

\subsection{Fractal Surfaces}

\begin{figure}[!ht]
\centering
\includegraphics[scale=0.8]{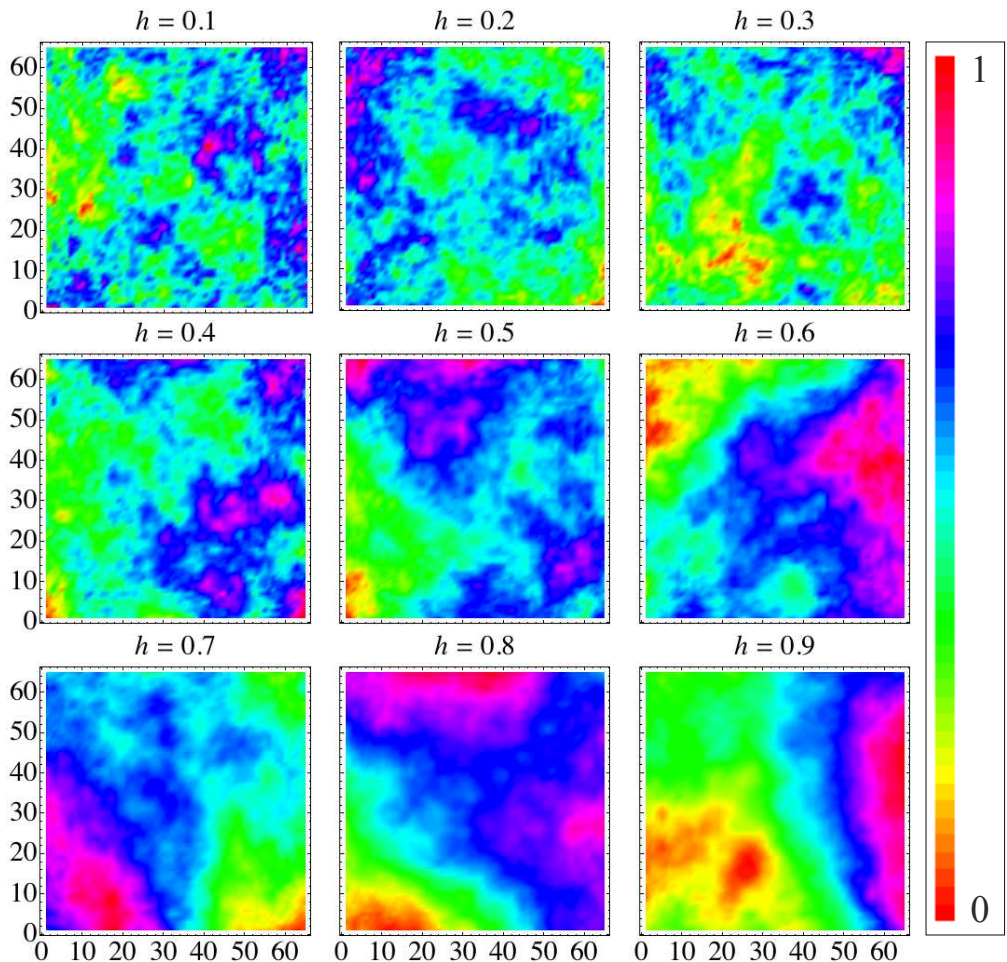}
\caption{\textbf{Examples of fractal surfaces obtained through the random midpoint displacement method.} 
These are $65\times 65$ surfaces ($k=6$) for different values of the Hurst exponent $h$. For easier
visualization, we have scaled the height of the surfaces in order to stay between $0$ and $1$.
{We note that for small values of $h$ the surfaces display an alternation of peaks and valleys 
(anti-persistent behavior) much more frequent than those one obtained for larger values of $h$.
For larger values of $h$, the surfaces are smoother reflecting the persistent behavior induced by the
value of $h>0.5$.} 
}\label{fig:midpoint}
\end{figure}

We generate fractal surfaces through the random midpoint displacement algorithm~\cite{Fournier}.
This algorithm starts with a square. For each vertex, we assign a random value 
representing the surface height. Next, we add a new point located at the 
center of the initial square. We set the height of this point equal to the average
height of the previous four vertex plus a Gaussian random number with zero mean
and standard-deviation $\delta_1$. We also add four points located at the middle
segments which connects each initial vertex. For these four points, the heights
are equal to the average value between the two closest vertex and the middle point 
plus a Gaussian random number with zero mean and standard-deviation $\delta_1$.
Now, we imagine that these 9 points represent four new squares and, for each one, 
we apply the previous procedure using $\delta_2$. By repeating this process $k$ 
times and using $\delta_k=\delta_0 \,2^{-\frac{k h}{2}}$, we should obtain a 
square surface of side $2^{k}+1$ with fractal properties. Here, $h$ is the Hurst 
exponent and $D=3-h$ is the surface fractal dimension. 
Figure \ref{fig:midpoint} shows several surfaces generated through this procedure 
for different values of $h$.

\begin{figure}[!ht]
\centering
\includegraphics[scale=0.6]{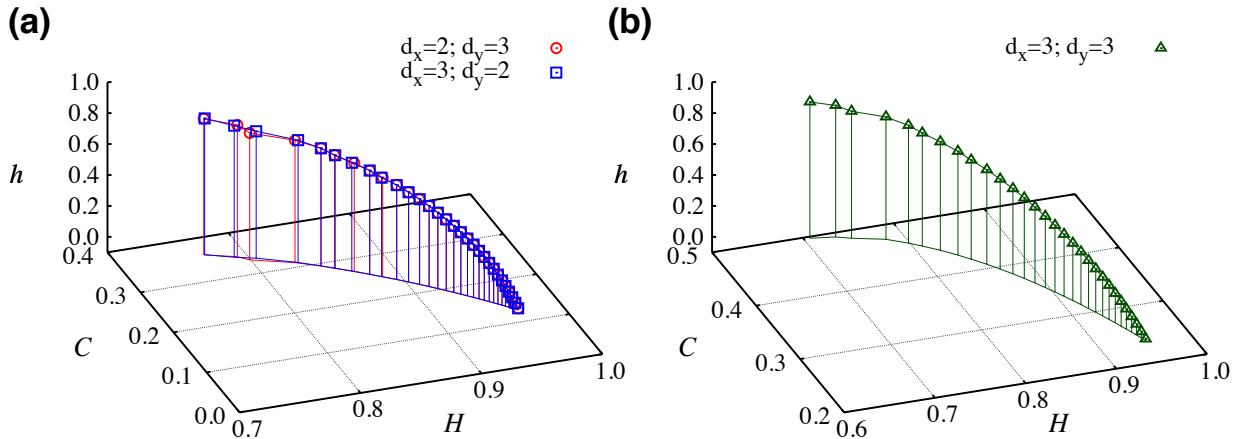}
\caption{\textbf{Dependence of the complexity-entropy causality plane on Hurst exponent $h$.} We have employed 
fractal surfaces of size $1025\times 1025$ ($k=10$). In (a) we plot $C$ and $H$ versus $h$ for the embedding 
dimensions $d_x=2$ and $d_y=3$ (circles) and also for $d_x=3$ and $d_y=2$ (squares). 
We note the invariance of the index against the rotation $d_x\to d_y$ and $d_y\to d_x$. In (b) we plot the diagram
for $d_x=d_y=3$. We observe changes in the scale of $C$ and $H$ caused by the increasing number of states.
{In both cases, as $h$ increases the complexity $C$ also increases while the permutation entropy $H$ decreases.
This behavior reflects the differences in the roughness shown in Fig.~\ref{fig:midpoint}. For values of $h<0.5$
the surface is anti-persistent which generates a flatter distribution for the values of $P=\{p(\pi)\}$ 
leading to values of $C$ and $H$ closer to the aleatory limit ($C\to0$ and $H\to1$). 
For values of $h>0.5$ there is a persistent behavior in the surfaces heights 
which generates a more intricate distribution of $P=\{p(\pi)\}$ and, consequently, 
values of $H$ and $C$ that are closer to the middle of the causality plane (region of higher complexity).}
}\label{fig:midpointa}
\end{figure}

We apply our method for these surfaces aiming to verify how the permutation quantifiers $H$ and $C$
change with the Hurst exponent $h$, as it is shown in Fig.~\ref{fig:midpointa}. In these 3d plots,
we show the localization in the causality plane obtained for different values of $h$ evaluated from $1025\times1025$ surfaces 
($k=10$). In Fig.~\ref{fig:midpointa}a, we use $d_x=2$ and $d_y=3$ (circles), and $d_x=3$ and $d_y=2$
(squares) as embedding dimensions. Note that the values of $H$ and $C$ are practically invariant
under the rotation $d_x\to d_y$ and $d_y\to d_x$. This invariance is related to the fact that in these
fractal surfaces there is not preferential direction. In Fig.~\ref{fig:midpointa}b, we employ 
$d_x=3$ and $d_y=3$. We note basically the same dependence but a different range for $H$ and $C$,
since this change increases the number of accessible states. These results show that our method properly
differentiates fractal surfaces concerning the Hurst exponent. {Moreover, we investigate the
robustness of the permutation quantifiers under several realizations of the random midpoint displacement
algorithm and the results show that both indexes are very stable. For example, the standard-deviation 
in the values of $H$ and $C$ are usually smaller than $10^{-4}$ when considering $k=10$.
}

\subsection{Liquid Crystal Textures}

Another interesting application is related to different patterns that a thin film of a liquid crystal exhibits.
These textures are obtained by observing a thin sample of liquid crystal
placed between two crossed polarizers in a microscope. The textures give useful information about
the macroscopic structure of the liquid crystal. For instance, different phases have different typical 
textures, and by tracking their evolution one can properly identify the phase transition. 

\begin{figure}[!ht]
\centering
\includegraphics[scale=0.74]{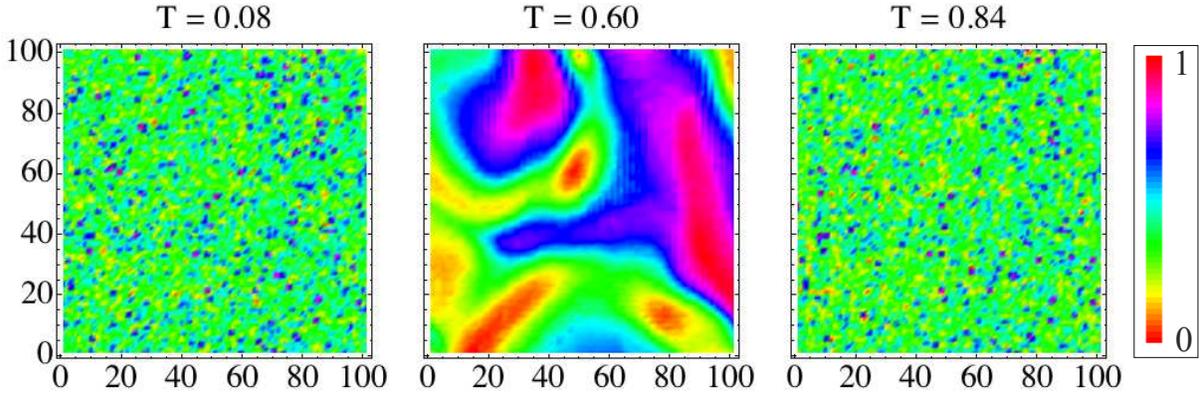}
\caption{\textbf{Characteristic textures of a lyotropic liquid crystal at different temperatures and phases.}
The lyotropic system used here is a mixture of potassium laurate $(\approx27.00\%)$, decanol $(\approx6.24\%)$ 
and deuterium oxide $(\approx66.76\%)$ --- suitable concentrations in order to get a isotropic $\to$ nematic $\to$ isotropic phase sequence~\cite{Saupe}. 
These images were constructed by observing the optical microscopy of a flat capillary which contains the mixture at different
temperatures. Here, we have used the average value of the pixels of the three layers (RGB) of the original image 
and a rescaled temperature.}\label{fig:lio}
\end{figure}

We first study a lyotropic liquid crystal under isotropic-nematic-isotropic phase transition. 
Figure~\ref{fig:lio} shows three snapshots of the texture at different temperatures. In
this case, we clearly note the differences in the textures. The leftmost and rightmost textures
are at the isotropic phase while the middle one is at the nematic phase. We observe that the pattern
is very complex for the nematic phase, while for the isotropic one it is basically random.

\begin{figure}[!ht]
\centering
\includegraphics[scale=0.72]{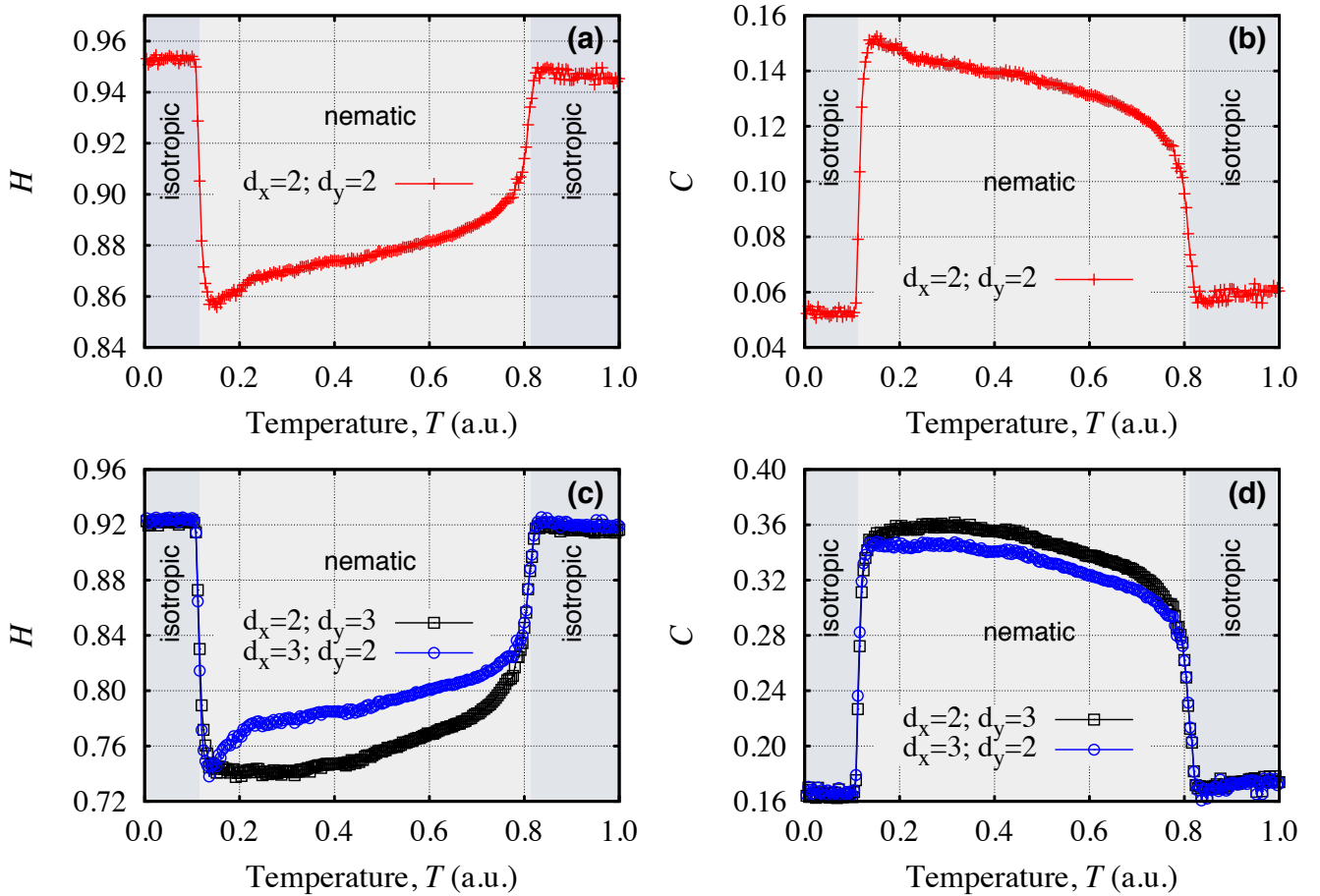}
\caption{\textbf{Dependence of the entropic indexes on the temperature of a lyotropic liquid crystal.}
We plot $H$ versus the temperature in (a) and $C$ versus the temperature in (b), where we employ $d_x=d_y=2$.
Figures (c) and (d) present the results for $d_x=2$ and $d_y=3$, and also for $d_x=3$ and $d_y=2$. The different
shaded areas represent the different liquid crystal phases. Note that the phase transitions are properly
identified in all cases. Due to the asymmetry of the elongated capillary tube where the liquid crystal sample is placed,
$H$ and $C$ present slight differences under the rotation $d_x\to d_y$ and $d_y\to d_x$.
}\label{fig:lioana}
\end{figure}

We calculate $H$ and $C$ as a function of the temperature for different values of the embedding
dimensions, as it is shown in Fig.~\ref{fig:lioana}. In these plots, the different shaded regions represent
the different liquid crystal phases. We note that the phase transitions are successful identified
independently of $d_x$ and $d_y$. However, Fig~\ref{fig:lioana}c and \ref{fig:lioana}d show a slight 
different dependence of $H$ and $C$ versus the temperature when considering $d_x=2$ and $d_y=3$ or
$d_x=3$ and $d_y=2$. Because the liquid crystal sample is placed in elongated capillary tube, there
is a surface effect that act on the liquid crystal molecules. This effect is usually amplified at the phase 
transition and it is also the reason for differences between the embedding dimensions. 

\begin{figure}[!ht]
\centering
\includegraphics[scale=0.53]{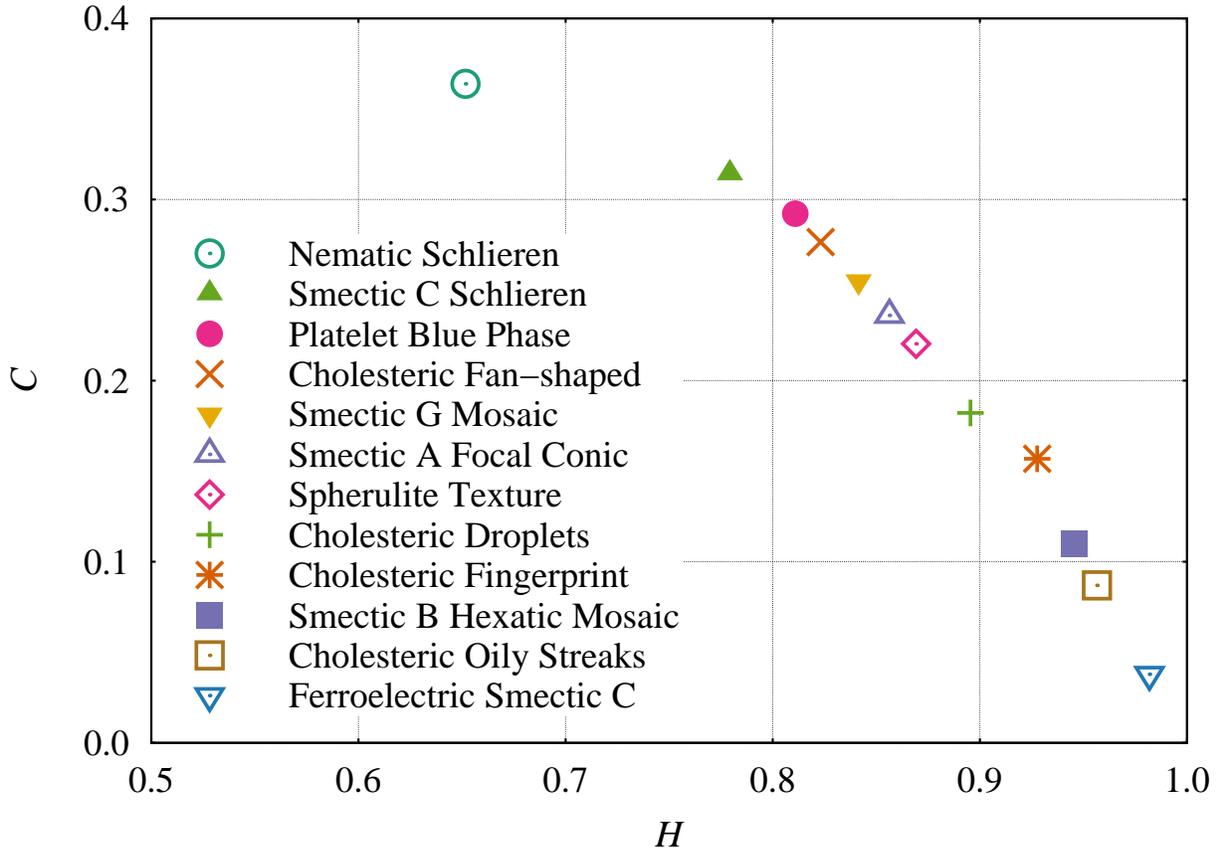}
\caption{\textbf{Complexity-entropy causality plane evaluated for several liquid crystal textures~\cite{Kent}}. 
Here, we have used the averaged pixel values of the three layers (RGB) of 
the original image and $d_x=2$ and $d_y=3$. The image sizes are about $270 \times 200$ pixels.
{We note that each texture has a unique position in the causality plane which indicates that the permutation quantifiers are capable of differentiate not only transitions involving the isotropic phase, but also smoother phase transitions. We further observe that some high ordered phase such as the blue phase are located at the central part
of the causality plane (region of higher complexity), while other phases which present a large number of defects such
as the Smectic B and C are closer to the aleatory limit ($C\to0$ and $H\to1$).
}
}\label{fig:lc-textures}
\end{figure}

In this particular phase transition, the difference between the textures are large enough that it can be
identified just by visual inspection. However, this is not the usual case and many phase transitions
are very difficult to identify. In this context, an interesting question is whether our method
can help to distinguish different phases. To address this question, we evaluate $H$ and $C$
for twelve characteristic textures of different liquid crystals. We download these textures from the
webpage of the Liquid Crystal Institute at Kent State University~\cite{Kent} and Fig.~\ref{fig:lc-textures}
shows the value of $H$ and $C$ for each texture in the causality plane. The results allow to conclude that the method 
ranks the textures in a kind of complexity order where each characteristic texture occupies a different 
place in this representation space. Moreover, the different values of  $H$ and $C$ indicate that the permutation
quantifiers can also identify smooth phase transitions. 

{Naturally, the location of each texture in the causality plane should be related to physical 
properties of the liquid crystals. A better understanding of the relation between the permutation
quantifiers and these physical attributes may deserves a more careful investigation since some
properties of liquid crystals such as the order parameter can be quite hard to empirically measure.
In this context, the existence of a clear relation between, for example, the order parameter and $H$ or
$C$ will be experimentally handy. Here, we just have the pictures of the textures in such a way that
is very hard to point out these relationships. However, a visual inspection of Fig.~\ref{fig:lc-textures}
suggests that some of the more ordered phases, such as the blue phase (this phase display a cubic 
structure of defects), are located in the central part of the causality plane (region of higher complexity),
while other textures which present a large number of non-ordered defects, such as the Smectic B and C, are positioned
closer to the aleatory limit ($C\to0$ and $H\to1$). Thus, it seems that the permutation quantifiers are
capturing in somehow the competition between the orientational order of the phase and, also, the number of
defects present in the textures.
}

\clearpage
\subsection{Ising Surfaces}
\begin{figure}[!ht]
\centering
\includegraphics[scale=0.77]{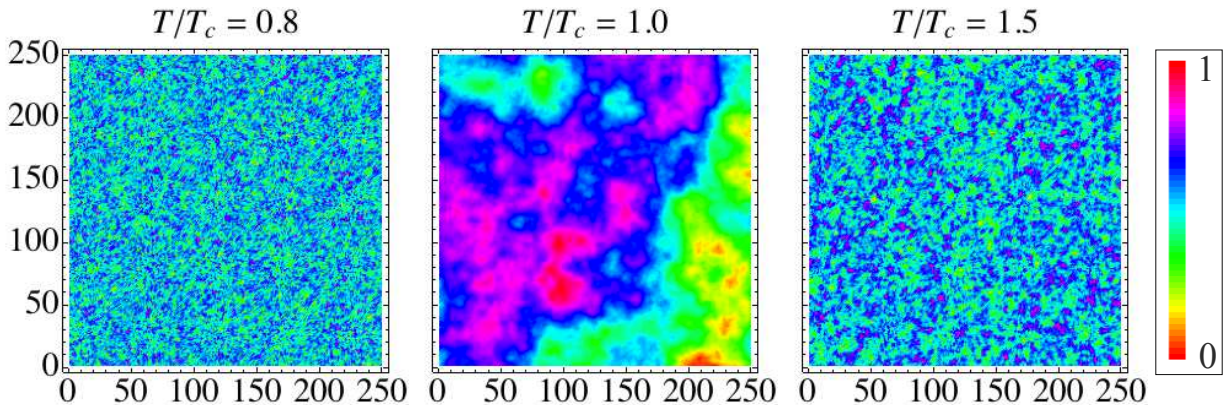}
\caption{\textbf{Examples of Ising surfaces for three different temperatures.} 
These surfaces were obtained after $10^5$ Monte Carlo steps for three different temperatures: 
below $T_c$, at $T_c$ and above $T_c$. In these plots, the height values were scaled to stay between $0$ and $1$.
{We note that for temperatures higher or lower than $T_c$, the surfaces exhibit an almost random pattern. For values of the temperature closer to $T_c$ the surfaces exhibit a more complex pattern, reflecting the long-range correlations
that appear among the spin sites during the phase transition.}
}\label{fig:isingsurf}
\end{figure}

As a last application, we study the permutation measures $H$ and $C$ applied to Ising surfaces~\cite{Brito,Brito2}.
These surfaces are obtained by accumulating the lattice spin values $\sigma_i\in\{-1,1\}$ of the Ising model
defined by the Hamiltonian
\begin{equation}
\mathcal H = -\sum_{\langle i,j \rangle} \sigma_i \sigma_j\,,
\end{equation}
where the sum is over all the pairs of first neighbor sites in the lattice. We numerically solve this spin-$1/2$ 
Ising model on a $L\times L$ lattice using the Monte Carlo method with periodic boundary conditions. By using the 
spin values, we define the surface height for each lattice site $i$ as 
\begin{equation}
\mathcal S_i = \sum_{t} \sigma_i(t)\,,
\end{equation}
where $t$ represents the number of Monte Carlo steps. In Fig.~\ref{fig:isingsurf}, we show three surfaces 
obtained though this procedure for different values of the reduced temperature $T/T_c$, where $T_c=2/\ln(1+\sqrt{2})$ is the
critical temperature of the model. We note the complex pattern exhibited by the surface for $T/T_c = 1$, and the almost random
patterns for $T/T_c>1$ and $T/T_c<1$. 

\begin{figure}[!ht]
\centering
\includegraphics[scale=0.7]{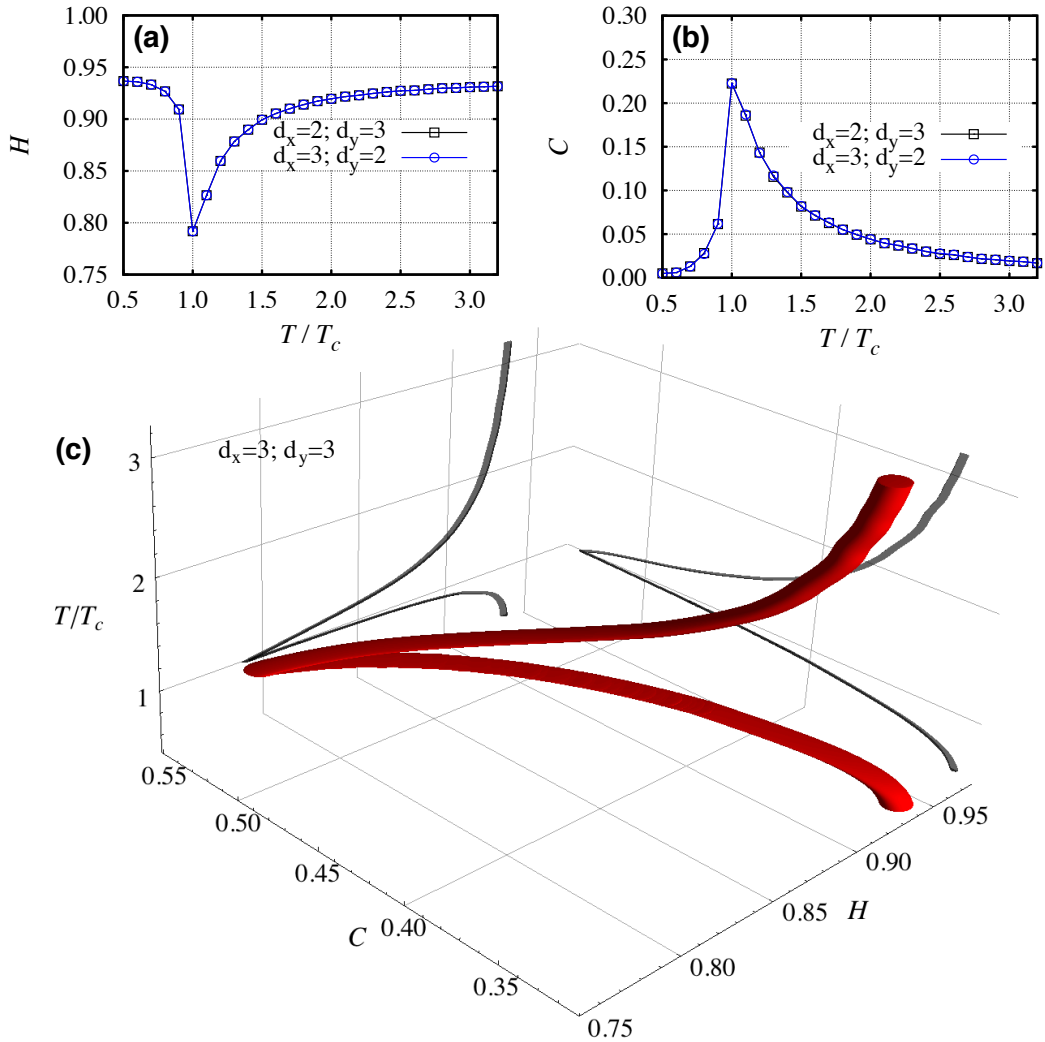}
\caption{\textbf{Dependence of the entropic indexes on the reduced temperature for Ising surfaces.} (a) The permutation
entropy $H$ and (b) the complexity measure $C$ versus the reduced temperature for $d_x=2$ and $d_y=3$, and also for $d_x=3$ 
and $d_y=2$. We note invariance of indexes under the rotation $d_x\to d_y$ and $d_y\to d_x$. 
(c) A 3d visualization of the Ising model phase transition when considering $d_x=d_y=3$. The gray shadows represent the dependences
of $T/T_c$ on $H$ and of $T/T_c$ on $C$.
}\label{fig:isingsurftemp}
\end{figure}

We first investigate the dependence of $H$ and $C$ on the reduced temperature $T/T_{C}$ after a large number of Monte Carlo steps 
($10^5$) and for $L=500$. Figures~\ref{fig:isingsurftemp}a and \ref{fig:isingsurftemp}b show $H$ and $C$ for $d_x=2$ and $d_y=3$, and
for the rotation $d_x\to d_y$ and $d_y\to d_x$. We note that, at the critical temperature, both indexes display a sharp peak and that they
are invariant under the rotation. Moreover, Fig.~\ref{fig:isingsurftemp}c presents a 3d visualization of the phase transition
for $d_x=d_y=3$. This higher-dimensional representation can be useful when investigating more complex phase transitions, since
a greater number of degrees of freedom allows the critical point to be more visible. 

\begin{figure}[!ht]
\centering
\includegraphics[scale=0.61]{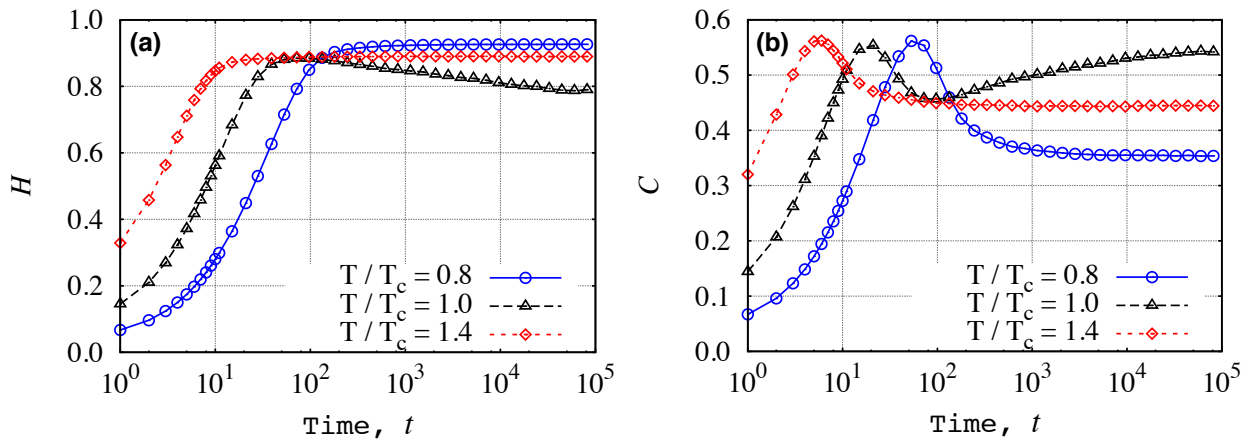}
\caption{\textbf{Dependence of the entropic indexes on the number of Monte Carlo steps.} Here, $t$ denotes
the number of Monte Carlo steps and the reduced temperatures are indicated in the plots. 
In (a) we show $H$ versus $t$ and in (b) $C$ versus $t$ for $d_x=d_y=3$. 
We note the stability of both indexes after $\sim10^4$ Monte Carlo steps.}\label{fig:isingsurfttime}
\end{figure}

We further study the temporal evolution of $H$ and $C$ for different reduced temperatures, as it is shown in Fig.~\ref{fig:isingsurfttime}.
The initial values of the spins were chosen equal to $1$ and, as we can see, the values for $H$ and $C$ are different just 
after one Monte Carlo step. For $T\neq T_c$, the value of $H$ increases over time and around $t\sim10^2$ it reaches a 
plateau. For $T=T_c$, the value of $H$ increases up to a maximum value around $t\sim10^2$ and then starts to approach 
a lower plateau value. A striking behavior is observed for $C$, where for all temperatures the complexity displays
a maximum value before it begins to approach a plateau value. It is worth noting that both quantifiers are 
very stable after $\sim10^4$ Monte Carlo steps.  

\clearpage 
\section{Summary}

We have proposed a generalization of the complexity-entropy causality plane to higher-dimensional patterns. We applied 
this approach  to fractal surfaces, liquid crystal textures and Ising surfaces. It was shown that the indexes $H$
and $C$ performed very well for distinguishing between the different roughness of the fractal surfaces. The indexes properly 
identified the phase transitions of a lyotropic liquid crystal and sorted different characteristic textures in a kind
of complexity order. Finally, concerning the Ising surfaces, the indexes not only had identified the critical temperature,
but also proved to be stable after $\sim 10^4$ Monte Carlo steps. The method also has a very fast and simple 
numerical evaluation. Taking into account all these findings, we are very optimist that our method can reduce the gap between one-dimensional 
complexity measures and the higher-dimensional ones.

\section{Acknowledgments}
HVR, RSM and EKL are grateful to CNPq and CAPES (Brazilian agencies) for the financial support. HVR also thanks 
CAPES for financial support under the process No.~5678-11-0. LZ was supported by Consejo Nacional de 
Investigaciones Cient\'ificas y T\'ecnicas (CONICET), Argentina.


\end{document}